\documentclass[final]{siamltex}
\usepackage[parfill]{parskip}    
\usepackage{amsmath,amssymb,latexsym,enumerate,mathrsfs}
\usepackage{hyperref}
\usepackage{array}
\usepackage{times}
\usepackage[]{mdframed}
\usepackage{framed}
\usepackage{subfig}
\usepackage{graphicx}
\usepackage{hyperref} 
\usepackage{multicol}
\usepackage{framed}
\usepackage{moreverb}
\usepackage{marvosym}
\usepackage{color,soul}
\definecolor{lightblue}{rgb}{.90,.95,1}
\sethlcolor{yellow}

\usepackage{epstopdf}
\DeclareMathOperator*{\argminmax}{arg\,min\,max}

\newtheorem{assumption}{\bf Assumption}

\title{Simulation-based parameter estimation via a combination of embedded normalizing flows and implied empirical probabilities under moment restrictions}

\author{Getachew K. Befekadu\footnote{\scriptsize Getachew K. Befekadu is with the Department of Electrical \& Computer Engineering, College of Engineering, Physics, and Computing, The Catholic University of America, ~ Washington, DC 20064, USA. E-mail:\,{\tt befekadu@cua.edu}}}

\begin{document}
\maketitle

\renewcommand{\thefootnote}{\arabic{footnote}}

\begin{abstract}
In this work, we present a simulation-based parameter estimation framework for a model defined by a computational simulation of a physical system. Here, we specifically outline an estimation framework consisting of two closely-integrated steps, but with different instantiated tasks, that facilitate an overall end-to-end parameter estimation scheme. The first step involves utilizing an embedded normalizing flow (i.e., a parameterized bijective transformation) which is used to transform the unknown complex probability distribution of the residual information (i.e., information associated with the discrepancies between the physical observed data and the simulation outputs) into a simple base probability distribution corresponding to the transformed residual information. In the second step, an empirical-likelihood estimator methods, under moment restrictions, is utilized for imposing an indirect constrain on the base distribution, where such an instantiated task reasonably allows us to treat the transformed residual information as random variables arising from discretely distribution population with each transformed data point as a single-cell from a set of finite-cell contingencies. Moreover, we use first-order gradient methods for updating the estimated parameter values of the model defined by the computational simulation and the corresponding parametrized embedded normalizing flow, that call for all gradient-related information by leveraging implicitly differentiations of the empirical-likelihood function, while the later is constructed from the implied empirical probabilities under moment restrictions. Here, it is worth mentioning that the problem formulation presented in this work, which also highlights an information-theoretic interpretation, allows us to present a computational framework for algorithmic implementations. Finally, as a-by-product, the inverse of the parametrized embedded normalizing flow, w.r.t. the estimated (sub)-optimal parameter values, serves as a surrogate model for the computational simulation model, which provides useful information for quantifying model discrepancies and sensitivity analysis.

\end{abstract}
\begin{keywords} 
Bijective transformation, empirical-likelihood, gradient methods, minimax problem, moment restrictions, normalizing flow, optimization, parameter estimation, simulation model.

\end{keywords}
\begin{AMS}
 37C05, 37M05, 62D20, 62F10, 62G30, 90C46, 90C47
\end{AMS}

\section{Introduction} \label{S1}
In this work, we present a simulation-based parameter estimation framework for a model defined by a computational simulation of a physical system. Here, our main interest is to outline an estimation framework consisting of two closely-integrated steps, but with different instantiated tasks supported by their respective viable mathematical arguments, that facilitate an overall end-to-end parameter estimation scheme. The first step involves utilizing an embedded normalizing flow (i.e., a parameterized bijective transformation) which is used to transform adaptively the unknown complex probability distribution of the residual information (i.e., information associated with the discrepancies between the physical observed data and the simulation output data) into a simple base probability distribution corresponding to the transformed residual information. Whereas, in the second step, an empirical-likelihood estimator method under moment restrictions is utilized for imposing an indirect constrain on the base probability distribution, where such an instantiated task reasonably enough allows us to treat the transformed residual information as random variables arising from discretely distribution population with each transformed data point as a single-cell from a set of finite-cell contingencies.

Moreover, we use first-order gradient methods for updating the estimated parameter values of the model defined by the computational simulation and the corresponding parametrized embedded normalizing flow, that call for all gradient-related information, by leveraging implicitly differentiations of the empirical-likelihood function, where the later is constructed from the implied empirical probabilities under moment restrictions. Here, it is worth mentioning that the problem formulation presented in this work, which highlights an information-theoretic interpretation, allows us to provide a computational framework for algorithmic implementations.\footnote{In this work, our intent is to provide an end-to-end parameter estimation framework that evidently provides useful information for quantifying model discrepancies, sensitivity analysis or uncertainty quantification on quantities of interest (e.g., see \cite{r1}-\cite{r5} for interesting studies discussing the need of addressing such challenges), rather than considering a specific numerical problem or application.\label{Footnote-1}} Finally, as a-by-product, the inverse of the parametrized embedded normalizing flow, w.r.t. the estimated parameter values, serves as a surrogate model for the computational simulation model, that provides useful information for quantifying model discrepancies and performing sensitivity analysis.

The remainder of this work is organized as follows. In Section~\ref{S2}, we provide a formal problem statement along with core concepts and some general assumptions that are relevant for the exposition of the simulation-based parameter estimation framework. Here, we outlines two closely-integrated steps, supported by their respective viable mathematical arguments, facilitating an overall end-to-end parameter estimation framework. In Section~\ref{S3}, we present our main results, i.e., we present a computational framework that highlights an information-theoretic interpretation -- based-on the first-order gradient methods -- for updating the estimated parameter values corresponding to the model defined by the computational simulation and that of the parametrized embedded normalizing flow, which basically requires us to collect all gradient-related information by leveraging implicitly differentiations of the empirical-likelihood, where the later is constructed from the implied empirical probabilities under moment restrictions. 

\section{Background and preliminaries} \label{S2} 
In this section, we provide a formal problem statement along with core concepts and some general assumptions that are relevant for the exposition of our simulation-based parameter estimation framework. In particular, our estimation framework, which leverages the use of embedded normalizing flows and empirical-likelihood estimator method, under moment restrictions, consisting of the following core concepts and assumptions.
\begin{enumerate} [(a).]
\item {\it Physical observed data $y^{\rm obs}(x)$, with support of $\mathbf{Y} \subset \mathbb{R}^{d_y}$}: We assume that we are given a set of independent observed data from a physical system or real experimental process $\zeta(x)$ under certain initial conditions or control input settings of $x$, i.e., $y^{\rm obs}(x) = \bigl\{y^{\rm obs}(x_i)\bigr\}_{i=1}^{n}$ under different initial conditions or control input settings of $x$, where $y^{\rm obs}(x_i) = \zeta(x_i) + \epsilon_i$, for $i=1$, $2$ \ldots, $n$, and $\epsilon_i$ represents the measurement noise in the $i^{\rm th}$ observation. Here, we loosely characterize the nature of $\epsilon$, but we will assume that it belongs to a certain class of continuous distributions.
\item {\it Computational simulation output $y^{\rm sim}(x;\theta)$, with support of $\mathbf{Y} \subset \mathbb{R}^{d_y}$}: We assume that the model defined by computational simulation for the physical system is denoted by $\eta(x, \theta)$ with input vectors $(x,\theta) \in \mathbf{X} \times \Theta$, where $\mathbf{X} \subset \mathbb{R}^{d_x}$ and $\Theta \subset \mathbb{R}^{d_{\theta}}$. Note that $\theta$ denotes the unknown parameters in the model defined by a computational simulation, whereas $x$ represents the initial conditions or control input settings much the same as to those in the above real physical system. Moreover, at different initial conditions or input control settings of $x$, with a certain fixed parameter value for $\theta$, the computational simulation is assumed to provide a set of simulation outputs $y^{\rm sim}(x; \theta) = \bigl\{y^{\rm sim}(x_i; \theta) \bigr\}_{i=1}^{n}$, where $y^{\rm sim}(x_i;\theta) = \eta(x_i, \theta) + \delta(x_i)$, for $i=1$, $2$ \ldots, $n$, while $\delta(x)$ represents the discrepancy between the model defined by the computational simulation and the real physical system. Here, we follow a standard assumption, where the real physical system $\zeta(x)$ does not depend on the unknown parameter $\theta$, but we wish to estimate the true or best parameter value $\theta^{\ast}$ consistent with the physical observed data $y^{\rm obs}(x)$. Furthermore, we assume here that the computational simulation is fast enough and any queries about the computational simulation can be made at will in the overall end-to-end estimation framework.
\item {\it Embedded normalizing flow $\operatorname{T}_{\phi} (\cdot)$}: We consider a normalizing flow (i.e., an embedded bijective transformation which is parameterized by a parameter $\phi \in \Phi \subset \mathbb{R}^{d_{\phi}}$) for transforming the unknown complex probability distribution $p_{\rm r}(r)$ of the residual information $r(x; \theta) = \bigl\{r(x_i; \theta)\bigr\}_{i=1}^n$, with $r(x_i; \theta)= y^{\rm obs}(x_i) - y^{\rm sim}(x_i; \theta)$, for $i=1$, $2$, \dots, $^n$, (i.e., the discrepancies between the physical observed data $y^{\rm obs}(x)$ and the computational simulation output $y^{\rm sim,\theta}(x)$) into a simple base probability distribution $p_{\rm z}(z)$ corresponding to the transformed residual information $z\bigl(x; \theta, \phi\bigr)$, with support of $\mathbf{Z} \subset \mathbb{R}^{d_z}$ (and assuming that $d_z=d_y$), i.e., $z\bigl(x; \theta, \phi\bigr) = \bigl\{z\bigl(x_i; \theta, \phi\bigr)\bigr\}_{i=1}^{n}$, where $z_i\bigl(x_i; \theta, \phi\bigr) = \operatorname{T}_{\phi} \bigl(r(x_i; \theta)\bigr)$, for $i=1$, $2$ \ldots, $n$. Here, we assume that the transformed residual information $z\bigl(x; \theta, \phi\bigr)$ obtained from the embedded normalizing flow $\operatorname{T}_{\phi}(\cdot)$ provides systematic discrepancies that will play an important role during updating the parameter values for $(\theta, \phi)$ that are associated with the model defined by the computational simulation and the corresponding parametrized embedded normalizing flow.
\item {\it Empirical-likelihood estimator under moment restrictions}: We consider an empirical-likelihood estimator method under moment restrictions for the transformed residual information $z\bigl(x; \theta, \phi\bigr) = \bigl\{z\bigl(x_i; \theta, \phi\bigr)\bigr\}_{i=1}^{n}$, with some input vectors $(\theta, \phi) \in \Theta \times \Phi$ and the initial conditions or control input settings $x \in \mathbf{X}$. Here, the reasoning behind this instantiated step is to impose an indirect constraint on the base distribution $p_{\rm z}(z)$ corresponding to the transformed residual information $z\bigl(x; \theta, \phi\bigr)$, while, at the same time, such an instantiated step reasonably allows us to treat the transformed residual information as random variables arising from discretely distribution population with each transformed data point $z\bigl(x_i; \theta, \phi\bigr)$, for $i \in \{1,2, \ldots, n\}$, as a single-cell from a set of finite-cell contingencies. Moreover, the dual solvability conditions for the empirical-likelihood estimator, under moment restrictions, will coincide with that of the saddle-point optimality conditions for a certain {\it minmax} optimization problem, where such an instantiated step, by leveraging implicitly differentiations, will further allow us to collect all gradient-related information from the empirical-likelihood function, which is constructed from the implied empirical probabilities under moment restrictions. 
\item {\it First-order gradient methods}: In this work, we make use of first-order gradient methods in the overall end-to-end parameter estimation framework, where such methods are often the preferred methods for large-scale applications that require us to collect all gradient-related information by leveraging implicitly differentiations from the empirical-likelihood function.
\end{enumerate}
In what follows, w.r.t. the transformed residual information $z\bigl(x; \tilde{\theta}, \tilde{\phi}\bigr) = \operatorname{T}_{\tilde{\phi}} \bigl(r(x; \tilde{\theta})\bigr)$ and for some fixed parameter values of $(\tilde{\theta}, \tilde{\phi}) \in \Theta \times \Phi$, we wish to determine a vector of unknown parameters $\beta \in \mathbf{B}$ (with compact set $\mathbf{B} \subset \mathbb{R}^{d_{\beta}}$) satisfying the following moment restrictions 
\begin{align}
E\bigl(g\bigl(z\bigl(x; \tilde{\theta}, \tilde{\phi}\bigr), \beta\bigr)\bigr) &= 0, \label{Eq1}
\end{align}
where $g \colon \mathbb{R}^{d_z} \times \mathbb{B} \to \mathbb{R}^{d_g}$ is a measurable function for modeling moment restrictions, while $g\bigl(z\bigl(x; \tilde{\theta}, \tilde{\phi}\bigr), \beta\bigr)$, with support of $\mathbb{R}^{d_g}$, is typically interpreted as a vector of residuals. Note that the above problem statement is basically equivalent to the following empirical-likelihood estimator method, under moment restrictions, i.e.,
\begin{align}
\max_{\pi_i, 1 \le i \le n; \, \beta \in \mathbb{B}} &\, \sum\nolimits_{i=1}^n \log \pi_i, \label{Eq2}\\
\text{s.t.} & \notag\\
&  \sum\nolimits_{i=1}^{n} \pi_i g\bigl(z(x_i; \tilde{\theta}, \tilde{\phi}), \beta\bigr)\bigr) = 0, \notag\\
 & \sum\nolimits_{i=1}^{n} \pi_i = 1, \notag
\end{align}
where $\pi_i \ge 0$, for $i = 1$, $2$, \ldots, $n$, are the implied empirical probabilities associated with the transformed residual information $z\bigl(x; \tilde{\theta}, \tilde{\phi}\bigr)$ under the moment restrictions of Equation~\eqref{Eq1} (e.g., see \cite{r6}-\cite{r10} for additional discussions).

Here, we remark that the empirical likelihood estimator method, under moment restrictions, reasonably allows us to treat the transformed residual information as random variables arising from discretely distribution population with each transformed data point as a single-cell from a set of finite-cell contingencies. Moreover, this line of reasoning allows us to impose an indirectly constraint on the nature of the base distribution corresponding to the transformed residual information.

Throughout this paper, we assume the following statements necessary for the solvability of the optimization problem in Equation~\eqref{Eq2} under moment restrictions (e.g., see \cite{r9} for additional discussions).
\begin{assumption}\label{AS1}
\begin{enumerate} [(1).]
\item There exists a unique solution $\beta_{0} \in \operatorname{Int}(\mathbf{B})$ satisfying $E\bigl(g\bigl(z\bigl(x; \tilde{\theta}, \tilde{\phi}\bigr), \beta_{0}\bigr)\bigr) = 0 $, for $z\bigl(x; \tilde{\theta}, \tilde{\phi}\bigr)$, with support of $\mathbf{Z}$, and for some fixed parameter values of $(\tilde{\theta}, \tilde{\phi}) \in \Theta \times \Phi$.
\item The function $g\bigl(z\bigl(x; \tilde{\theta}, \tilde{\phi}\bigr), \beta \bigr)$ is continuous at each $\beta \in \mathbf{B}$ with probability one. Moreover, the following conditions hold true:
\begin{enumerate} [$~~~$(i).]
\item $E \bigl(sup_{\beta \in \mathbb{B}} \bigl\Vert g\bigl(z\bigl(x; \tilde{\theta}, \tilde{\phi}\bigr), \beta \bigr) \bigl \Vert^{\alpha} \bigr) < \infty$ for some $\alpha > 2$,
\item $E \bigl(sup_{\beta \in N_{\beta_{0}}} \bigl\Vert \nabla_{\beta} g\bigl(z\bigl(x; \tilde{\theta}, \tilde{\phi}\bigr), \beta \bigr) \bigl \Vert \bigr) < \infty$, where $N_{\beta_{0}} \subset \mathbf{B}$ denotes the neighborhood of $\beta_{0}$,
\item $\Omega = E \bigl( g\bigl(z\bigl(x; \tilde{\theta}, \tilde{\phi}\bigr), \beta\bigr) g\bigl(z\bigl(x; \tilde{\theta}, \tilde{\phi}\bigr), \beta\bigr)^T\bigr)\bigl\vert_{\beta= \beta_{0}}$ is a nonsingular matrix, and
\item$\operatorname{rank}\,E \bigl(\nabla_{\beta} g\bigl(z\bigl(x; \tilde{\theta}, \tilde{\phi}\bigr), \beta \bigr) \bigr) = d_{\beta}$,
\end{enumerate}
for $z\bigl(x; \tilde{\theta}, \tilde{\phi}\bigr)$, with support of $\mathbf{Z}$, and for some fixed parameter values of $(\tilde{\theta}, \tilde{\phi}) \in \Theta \times \Phi$.
\end{enumerate}
\end{assumption}

Moreover, we assume the following two assumptions hold true.
\begin{assumption}\label{AS2}
For some fixed parameter values of $(\theta, \phi) \in \Theta \times \Phi$, the normalizing flow $\operatorname{T}_{\phi}$ is a parametrized embedded bijective transformation, w.r.t. the parameter $\phi$, and further satisfies the following condition
\begin{align*}
p_{\rm r}\bigl(r(x;\theta, \phi)\bigr) &= p_{\rm z}\bigl(z(x;\theta, \phi)\bigr) \left\vert \frac{\partial z(x;\theta, \phi)}{\partial r(x;\theta, \phi)} \right \vert \\
                                &= p_{\rm z}\bigl(\operatorname{T}_{\phi} \bigl(r(x;\theta, \phi) \bigr)\bigr) \left\vert \operatorname{det} J_{\operatorname{T}_{\phi}} \bigl(r(x;\theta, \phi)\bigr) \right \vert,
\end{align*}
where $p_{\rm r}(r)$ represents the unknown complex probability distribution corresponding the residual information $r(x;\theta, \phi)$ (i.e., $r(x;\theta, \phi) \equiv y^{\rm obs}(x) - y^{\rm sim}(x; \theta)$, with support of $\mathbf{Y}$), while $p_{\rm z}(z)$ corresponds to the base probability distribution for the transformed residual information $z(x; \theta, \phi)$, with support of $\mathbf{Z}$. Moreover, $J_{\operatorname{T}_{\phi}}$ denotes a $d_{y} \times d_{y}$ Jacobian matrix corresponding to the normalizing flow $\operatorname{T}_{\phi}$.\footnote{Here, we remark that normalizing flows are a class of generative models for continuous random variables that provide a mechanism for defining expressive probability distributions (e.g., see \cite{r11} and \cite{r12} some interesting discussions in machine learning literature).} 
\end{assumption}

\begin{assumption}\label{AS3}
For some fixed parameter values of $(\theta, \phi) \in \Theta \times \Phi$, we assume that the empirical-likelihood estimator, under moment restrictions, in Equation~\ref{Eq2} will provide an indirect constraint on the base probability distribution $p_{\rm z}(z)$ corresponding to the transformed residual information $z(x;\theta,\phi)$, with support of $\mathbf{Z}$.
\end{assumption}

With a standard Lagrangian argument, the optimization problem in Equation~\eqref{Eq2} (i.e., the empirical likelihood estimator, under moment restrictions) can be further transformed to a non-constrained optimization problem as follows
\begin{align}
\mathscr{L}^{(\tilde{\theta}, \tilde{\phi})}\bigl(\beta, \pi, \lambda, \eta  \bigr) = \sum\nolimits_{i=1}^n \log \pi_i \, - \, & n \lambda^T\sum\nolimits_{i=1}^n \pi_i g\bigl(z\bigl(x_i; \tilde{\theta}, \tilde{\phi}\bigr), \beta \bigr) \notag \\ 
& \quad\quad\quad\quad - \eta \left(\sum\nolimits_{i=1}^n \pi_i - 1 \right), \label{Eq3}
\end{align}
where $\lambda \in \mathbb{R}^{d_g}$ and $\eta \in \mathbb{R}$ are the Lagrange multipliers associated with the constraints in the optimization problem of Equation~\eqref{Eq2}. Moreover, suppose that Assumptions~\ref{AS1} and ~\ref{AS2} hold true. Then, for a certain fixed value of $\hat{\beta} \in \operatorname{Int}(\mathbf{B})$, using the envelop theorem (e.g., see \cite{r13} and \cite{r14} for additional discussions), then it is easy to see that from the KKT's optimality conditions (e.g., see also \cite{r15}) that the implied empirical probabilities $\hat{\pi}_i\bigl(\hat{\beta}, \hat{\lambda}\bigl(\hat{\beta}\bigr)\bigr)$, for $i = 1$, $2$, \ldots, $n$, under moment restrictions, associated with the transformed residual information $z\bigl(x; \tilde{\theta}, \tilde{\phi}\bigr)$, satisfy the following primal and dual feasibility conditions
\begin{equation}
\text{Primal Feasibility:} \quad \quad \left\{ \begin{matrix} 
 \left(\sum\nolimits_{i=1}^{n} \hat{\pi}_i\bigl(\hat{\beta}, \hat{\lambda}\bigl(\hat{\beta}\bigr)\bigr) - 1 \right)= 0\vspace{1mm} \\
 n \sum\nolimits_{i=1}^{n} \hat{\pi}_i\bigl(\hat{\beta}, \hat{\lambda}\bigl(\hat{\beta}\bigr)\bigr) g\bigl(z\bigl(x_i; \tilde{\theta}, \tilde{\phi}\bigr), \hat{\beta}\bigr) = 0\vspace{2mm} 
\end{matrix}\right. \label{Eq4}
\end{equation}
and 
\begin{equation}
\text{Dual Feasibility:} \quad \left\{ \begin{matrix} 
\frac{1}{\hat{\pi}_i\bigl(\hat{\beta}, \hat{\lambda}\bigl(\hat{\beta}\bigr)\bigr)} - n \hat{\lambda}(\hat{\beta})^T g\bigl(z\bigl(x_i; \tilde{\theta}, \tilde{\phi}\bigr), \hat{\beta} \bigr) - \hat{\eta}(\hat{\beta}) = 0 \vspace{2mm} \\
 \quad\quad\quad\quad\quad\quad\quad\quad\quad\quad\quad ~~~ \text{for} \quad i = 1,2, \ldots, n \vspace{2mm} 
\end{matrix}\right. \label{Eq5}
\end{equation}
where $\hat{\lambda}(\hat{\beta})$ and $\hat{\eta}(\hat{\beta})$ are, respectively, the Lagrange multipliers associated with the constraints 
\begin{align*}
\sum\nolimits_{i=1}^n \hat{\pi}_i\bigl(\hat{\beta}, \hat{\lambda}\bigl(\hat{\beta}\bigr)\bigr) g\bigl(z\bigl(x_i; \tilde{\theta}, \tilde{\phi}\bigr), \hat{\beta} \bigr) = 0 \quad \text{and} \quad \sum\nolimits_{i=1}^n \hat{\pi}_i \bigl(\hat{\beta}, \hat{\lambda}\bigl(\hat{\beta}\bigr)\bigr) = 1.
\end{align*}

Note that, if we multiply Equation~\eqref{Eq5} on both sides by $\hat{\pi}_i \bigl(\hat{\beta},\hat{\lambda}\bigl(\hat{\beta}\bigr)\bigr)$ and sum overall $i$, i.e.,
\begin{align}
 \sum\nolimits_{i=1}^{n} \hat{\pi}_i \bigl(\hat{\beta},\hat{\lambda}\bigl(\hat{\beta}\bigr)\bigr) \left(\frac{1}{\hat{\pi}_i \bigl(\hat{\beta}, \hat{\lambda}\bigl(\hat{\beta}\bigr)\bigr)}  - n \hat{\lambda}(\hat{\beta})^T g\bigl(z\bigl(x_i; \tilde{\theta}, \tilde{\phi}\bigr), \hat{\beta} \bigr) - \hat{\eta}(\hat{\beta}) \right) = 0. \label{Eq6}
\end{align}
Then, we will have the following conditions, for $i=1$, $2$, \ldots, $n$, 
\begin{align*}
 \hat{\eta}(\hat{\beta})  = - n \quad \quad\text{and} \quad \quad \hat{\pi}_i \bigl(\hat{\beta}, \hat{\lambda}\bigl(\hat{\beta}\bigr)\bigr) = \frac{1}{n \bigl(1 + \hat{\lambda}(\hat{\beta})^T g\bigl(z\bigl(x_i; \tilde{\theta}, \tilde{\phi}\bigr), \hat{\beta} \bigr)\bigr)},
\end{align*}
that further implies the Lagrange multiplier $\hat{\lambda}(\tilde{\beta})$ satisfies the following condition
\begin{align}
\sum\nolimits_{i=1}^{n} \hat{\pi}_i  \bigl(\hat{\beta}, \hat{\lambda}\bigl(\hat{\beta}\bigr)\bigr) &= \sum\nolimits_{i=1}^{n}\frac{1}{n \bigl(1 +  \hat{\lambda}(\hat{\beta})^T g\bigl(z\bigl(x_i; \tilde{\theta}, \tilde{\phi}\bigr), \hat{\beta}\bigr)\bigr)} \notag \\
                                           & = 1, \label{Eq7}
\end{align}
i.e., implying the sum of the implied empirical probabilities, under moment restrictions, is unity. Moreover, for each $\hat{\beta} \in \operatorname{Int}(\mathbf{B})$, the Lagrange multiplier $\hat{\lambda}(\hat{\beta})$ also solves the following equation
\begin{align}
 \sum\nolimits_{i=1}^{n} \frac{g\bigl(z\bigl(x_i; \tilde{\theta}, \tilde{\phi}\bigr), \hat{\beta} \bigr)}{1 + \hat{\lambda}(\hat{\beta})^T g\bigl(z\bigl(x_i; \tilde{\theta}, \tilde{\phi}\bigr), \hat{\beta}\bigr)} = 0. \label{Eq8}
 \end{align}
Note that, when $\hat{\pi}_i\bigl(\hat{\beta}, \hat{\lambda}\bigl(\hat{\beta}\bigr)\bigr) g\bigl(z\bigl(x_i; \tilde{\theta}, \tilde{\phi}\bigr), \hat{\beta} \bigr)$ is uniformly small for all $i \in \{1,2, \ldots, n\}$, if the first-order KKT conditions for the dual variable $\hat{\lambda}(\hat{\beta})$ hold true. Then, the corresponding implied empirical probabilities, under moment restrictions, will further satisfy the following sample moment restrictions 
\begin{align*}
\sum\nolimits_{i=1}^{n} \hat{\pi}_i\bigl(\hat{\beta}, \hat{\lambda}\bigl(\hat{\beta}\bigr)\bigr)  g\bigl(z\bigl(x_i; \tilde{\theta}, \tilde{\phi}\bigr), \hat{\beta} \bigr) = 0.
\end{align*}
Moreover, the corresponding empirical log-likelihood function $\ell^{(\tilde{\theta}, \tilde{\phi})}\bigl(\hat{\beta}, \hat{\lambda}\bigl(\hat{\beta}\bigr)\bigr)$ can be recovered from the implied empirical probabilities, under moment restrictions, as follows:
\begin{align*}
\ell^{(\tilde{\theta}, \tilde{\phi})}\bigl(\hat{\beta}, \hat{\lambda}\bigl(\hat{\beta}\bigr)\bigr) &= \sum\nolimits_{i=1}^{n} \log \hat{\pi}_i\bigl(\hat{\beta}, \hat{\lambda}\bigl(\hat{\beta}\bigr)\bigr)\\
                                                        &= - n \log n - \sum\nolimits_{i=1}^{n} \log \left(1 + \hat{\lambda}(\hat{\beta})^T g\bigl(z\bigl(x_i; \tilde{\theta}, \tilde{\phi}\bigr), \hat{\beta} \bigr)\bigr)\right).
\end{align*}

In what follows, we establish a connection between the dual solvability conditions for the empirical-likelihood estimator, under moment restrictions, with that of the saddle-point optimality conditions for a certain {\it minmax} optimization problem. Note that, from Assumption~\ref{AS1}, if the function $g(\cdot, \beta)$ is continuously differentiable w.r.t. $\beta \in \mathbf{B}$. Then, for any values of $(\theta, \phi) \in \Theta \times \Phi$, the saddle-point optimality conditions for the following {\it minmax} optimization problem
\begin{align}
\bigl(\hat{\beta}, \hat{\lambda} (\hat{\beta}) \bigr) \in \argminmax_{~~\beta \in \mathbf{B},\, \lambda \in \mathbb{R}^{d_g}} \sum\nolimits_{i=1}^n \rho \left( \lambda^T  g\bigl(z\bigl(x_i; \theta, \phi\bigr), \beta\bigr)\right), \label{Eq9}
\end{align}
with $\rho(v) = \log(1+v)$, with $v \in \mathbf{V} \equiv (-1, \infty)$, will coincide with that of the dual feasibility conditions in Equation~\eqref{Eq5} (e.g., see \cite{r10} for some related discussions; see also \cite{r17} and \cite{r18} for general discussions on bilevel optimization problems), i.e.,
\begin{align}
\hat{\pi}_i \bigl(\hat{\beta}, \hat{\lambda}\bigl(\hat{\beta}\bigr)\bigr) = \frac{1}{n \left(\hat{\lambda}(\hat{\beta})^T g\bigl(z\bigl(x_i; \theta, \phi\bigr), \hat{\beta} \bigr) + 1\right)}, \quad i=1,2, \ldots, n. \label{Eq10}
\end{align}
 Moreover, assuming that $\hat{\beta}$ and $\hat{\lambda}\bigl(\hat{\beta}\bigr)$ continuously depend on the parameters $(\theta, \phi) \in \Theta \times \Phi$. With a slight abuse of notation, the empirical log-likelihood function $\ell \bigl(\hat{\beta}, \hat{\lambda}\bigl(\hat{\beta}\bigr)\bigr)$ can be recovered as follows
\begin{align}
\ell \bigl(\hat{\beta}, \hat{\lambda}\bigl(\hat{\beta}\bigr)\bigr) &= - \log n - \sum\nolimits_{i=1}^{n} \log \left(1 + \hat{\lambda}(\hat{\beta})^T g\bigl(z\bigl(x_i; \theta, \phi\bigr), \hat{\beta} \bigr)\bigr)\right). \label{Eq11}
\end{align}
In the following section, by leveraging implicitly differentiations through the above {\it minmax} optimization problem, we will establish a connection that allows us to collect all gradient-related information from the empirical-likelihood function, under moment restrictions, that are required for updating the parameter values associated with the model defined by the computational simulation and the parametrized embedded normalizing flow. 

\section{Main results} \label{S3}
In this section, we present our main results, i.e., we provide an over end-to-end parameter estimation framework for updating parameter values corresponding to the model defined by the computational simulation and that of the parametrized embedded normalizing flow. We specifically use first-order gradient methods that require us to collect all gradient-related information from the empirical-likelihood function, which is constructed from the implied empirical probabilities, under moment restrictions. 

\begin{figure}[h]
\begin{center}
 \includegraphics[scale=0.58]{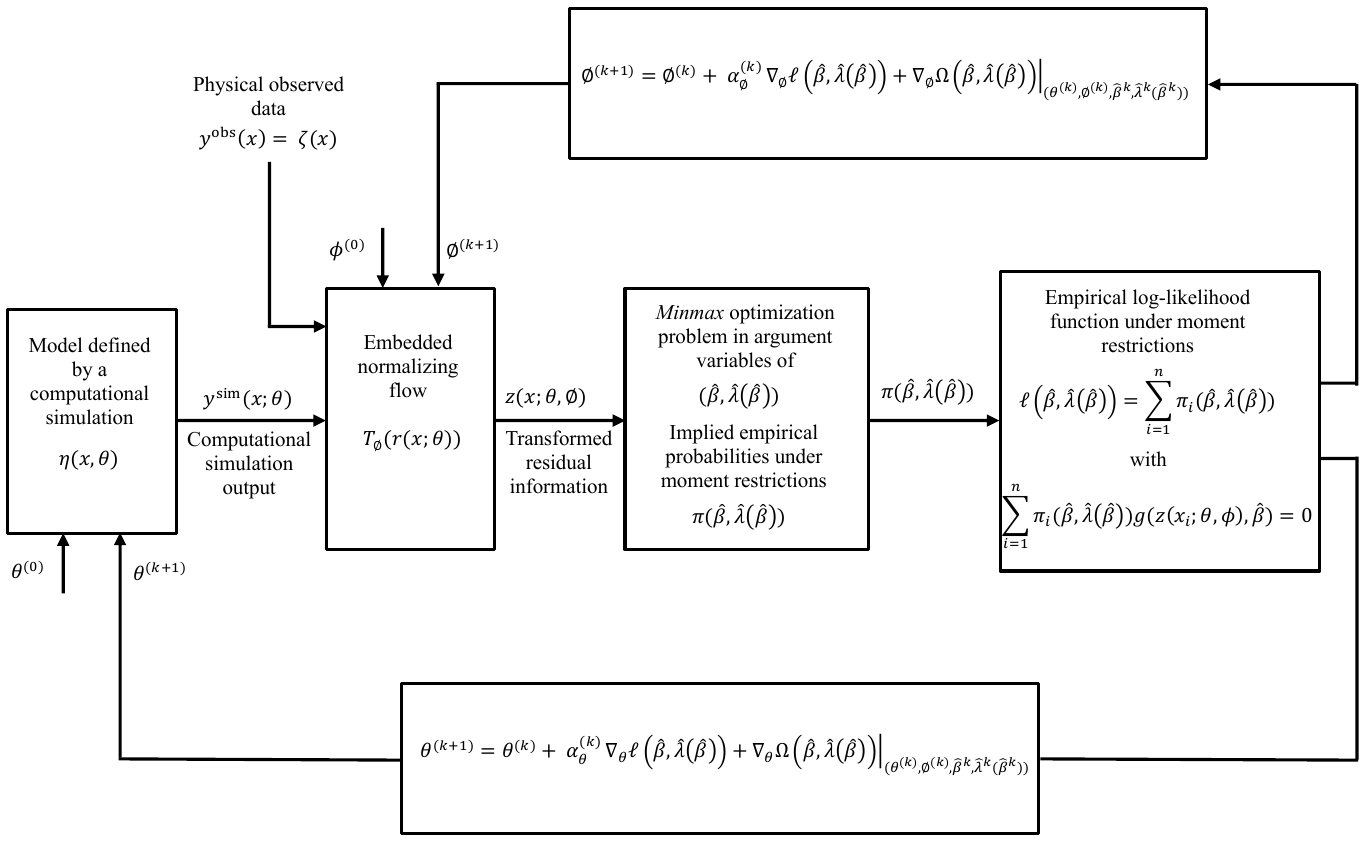}
 \caption{The overall, end-to-end, simulation-based parameter estimation scheme} \label{Figure1}
\end{center}
\end{figure}

Note that, for any fixed values of $(\theta, \phi) \in \Theta \times \Phi$, solving the {\it minmax} optimization problem in Equation~\eqref{Eq9}, based on the first-order gradient methods, requires all gradient-related information corresponding to the solution of the {\it inner maximization problem} w.r.t. that of the variable of the {\it outer minimization problem}. Recall that $\hat{\beta}$ and $\hat{\lambda}\bigl(\hat{\beta}\bigr)$ continuously depend on the parameters $(\theta, \phi) \in \Theta \times \Phi$ and, hence, computing all gradient-related information requires us to leverage the implicitly differentiations through the {\it minmax} optimization problem of Equation~\eqref{Eq9}. Here, we claim that the {\it inner maximization problem} is efficiently solvable and there exist a set of computational schemes for computing all gradient-related information at the current solutions. More precisely, we require the following gradient-related information
\begin{enumerate} [(i).]
\item $\nabla_{\theta} \ell\bigl(\hat{\beta}, \hat{\lambda}(\hat{\beta})\bigr)$,
\item $\nabla_{\phi} \ell\bigl(\hat{\beta}, \hat{\lambda}(\hat{\beta})\bigr)$,
\item $\nabla_{\theta} \Omega\bigl(\hat{\beta}, \hat{\lambda}(\hat{\beta})\bigr)$ and $\nabla_{\phi} \Omega\bigl(\hat{\beta}, \hat{\lambda}(\hat{\beta})\bigr)$, where $\Omega\bigl(\hat{\beta}, \hat{\lambda}(\hat{\beta})\bigr)$ is a regularization term. 
\end{enumerate} 
Note that the above gradient-related information further call for the following quantities: $\partial \ell\bigl(\hat{\beta}, \hat{\lambda}(\hat{\beta})\bigr) \bigl/ \partial \hat{\beta}$, $\partial \ell\bigl(\hat{\beta}, \hat{\lambda}(\hat{\beta})\bigr) \bigl/ \partial \hat{\lambda}(\hat{\beta})$, $\nabla_{\theta} \hat{\beta}$, $\nabla_{\phi} \hat{\beta}$, $\nabla_{\theta} \hat{\lambda}(\hat{\beta})\bigr)$, $\nabla_{\phi} \hat{\lambda}(\hat{\beta})\bigr)$ and $\nabla_{\hat{\beta}} g\bigl(\operatorname{T}_{\phi}(.), \hat{\beta}\bigr)$, that necessitate implicitly differentiations through the {\it minmax} optimization problem (and the computational simulation model) at the current solutions $\bigl(\theta, \phi, \hat{\beta}, \hat{\lambda}(\hat{\beta})\bigr) \in \Theta \times \Phi \times \mathbb{R}^{d_{\beta}} \times \mathbb{R}^{d_g}$. As a result of this, we will have the following overall computational schemes appropriate for updating the estimated parameter values (see also Figure~\ref{Figure1} that shows the overall end-to-end viable algorithmic implementation):
\begin{align*}
&\text{(i)}. \quad \quad \theta^{(k+1)} = \theta^{(k)} + \alpha_{\theta}^{(k)} \left(\nabla_{\theta} \ell\bigl(\hat{\beta}, \hat{\lambda}(\hat{\beta})\bigr) + \nabla_{\theta} \Omega\bigl(\hat{\beta}, \hat{\lambda}(\hat{\beta})\bigr)\right) \biggl \vert_{\bigl(\theta^{(k)}, \phi^{(k)}, \hat{\beta}^{(k)}, \hat{\lambda}^{(k)}(\hat{\beta}^{(k)}) \bigr)} \\ ~\\
&\text{(ii)}. \quad \quad \phi^{(k+1)} = \phi^{(k)} + \alpha_{\phi}^{(k)} \left(\nabla_{\phi} \ell\bigl(\hat{\beta}, \hat{\lambda}(\hat{\beta})\bigr) + \nabla_{\phi} \Omega\bigl(\hat{\beta}, \hat{\lambda}(\hat{\beta})\bigr)\right)\biggl \vert_{\bigl(\theta^{(k)}, \phi^{(k)}, \hat{\beta}^{(k)}, \hat{\lambda}^{(k)}(\hat{\beta}^{(k)}) \bigr)}
\end{align*}
where $\alpha_{\theta}^{(k)}$ and $\alpha_{\phi}^{(k)}$ are step-sizes.

Finally, it is worth remarking that, as a-by-product, the inverse of the parametrized embedded normalizing flow, w.r.t. the optimal estimated parameter values $\bigl(\theta^{\ast}, \phi^{\ast}, \hat{\beta}^{\ast}, \hat{\lambda}^{\ast}(\hat{\beta}^{\ast}) \bigr)$, serves as a surrogate model for the computational simulation model $\eta(x, \theta^{\ast})$, that provides useful information for quantifying model discrepancies or sensitivity analysis. For example, consider the following input set $\mathbf{M} = \left\{ (x,\theta^{\ast}) \in (\mathbf{X}, \Theta) ~ \bigl \vert ~ \eta(x,\theta^{\ast}) + \delta(x) = \zeta(x) \right \}$, with restriction of $y^{\rm obs}(x) - y^{\rm sim}(x; \theta^{\ast}) = \operatorname{T}_{\phi^{\ast}}^{-1} \bigl(z\bigl(x; \theta^{\ast},\phi^{\ast} \bigr)\bigr)$ and $\phi^{\ast} \in \Phi$. Moreover, if we define the $\alpha$-quantile, w.r.t. the implied empirical probabilities $\hat{\pi}_i^{\ast}$, for $i=1$, $2$, \ldots, $n$, under moment restrictions $\sum\nolimits_{i=1}^{n} \hat{\pi}_i^{\ast}\bigl(\hat{\beta}^{\ast}, \hat{\lambda}^{\ast}\bigl(\hat{\beta}^{\ast}\bigr)g\bigl(z(x_i; \theta^{\ast}, \phi^{\ast}), \hat{\beta}\bigr) = 0$, at support points $\bigl\{z(x_i; \theta^{\ast},\phi^{\ast})\bigr\}_i^n$, as
\begin{align*}
\operatorname{Quantile}_{\alpha} \left(\sum_{i=1}^{n} \hat{\pi}_i^{\ast}\bigl(\hat{\beta}^{\ast}, \hat{\lambda}^{\ast}\bigl(\hat{\beta}^{\ast}\bigr)\bigr) \delta_{z(x_i; \theta^{\ast},\phi^{\ast})}\right) = \inf \left\{ z ~ \biggl \vert ~ \sum_{\substack{i=1 \\ z(x_i; \theta^{\ast},\phi^{\ast})\le z}}^{n} \hat{\pi}_i^{\ast}\bigl(\hat{\beta}^{\ast}, \hat{\lambda}^{\ast}\bigl(\hat{\beta}^{\ast}\bigr)\bigr) \ge \alpha \right\},
\end{align*}
where $\delta_{z(x_i; \theta^{\ast},\phi^{\ast})}$ is a unit point mass at $z(x_i; \theta^{\ast},\phi^{\ast})$, for $i \in \{1,\,2, \ldots, n\}$. Then, this will allow us to construct a confidence set with a certain miscoverage level of $\alpha \in (0,1)$ for some quantities of interest.

\subsection*{Information-theoretic Interpretations}: In this subsection, we highlight an information-theoretic interpretation, in which we further provide the rationale behind our overall, end-to-end, simulation-based parameter estimation framework. Recall that, for some fixed values $(\theta, \phi) \in \Theta \times \Phi$, the estimated parameter $\hat{\beta} \in \operatorname{Int}(\mathbf{B})$ based on the empirical likelihood estimator, under moment restrictions, satisfies the following sample moment restrictions
\begin{align*}
\sum\nolimits_{i=1}^n \hat{\pi}_{i}\bigl(\hat{\beta}, \hat{\lambda}(\hat{\beta})\bigr) g\bigl(z(x_i;  \theta, \phi), \hat{\beta}\bigr) = 0,
\end{align*}
where the corresponding implied empirical probabilities are assumed to be strictly positive and their sum is unity, i.e., $\sum\nolimits_{i=1}^n \hat{\pi}_{i}\bigl(\hat{\beta}, \hat{\lambda}(\hat{\beta})\bigr)  = 1$, with $\hat{\pi}_{i}\bigl(\hat{\beta}, \hat{\lambda}(\hat{\beta})\bigr) > 0$, for $i=1$, $2$, \ldots, $n$. 
Moreover, we consider the following total variational distance,
\begin{align}
\bigl\Vert \pi^{(0)} - \hat{\pi} \bigl(\hat{\beta}, \hat{\lambda}(\hat{\beta})\bigr) \bigr \Vert_{\rm TV} = \sum\nolimits_{i=1}^n \left \vert (1/n) - \hat{\pi}_i \bigl(\hat{\beta}, \hat{\lambda}(\hat{\beta})\bigr)\right\vert,  \label{Eq12}
\end{align}
i.e., a discrepancy gap measure between the implied empirical probabilities $\hat{\pi} \bigl(\hat{\beta}, \hat{\lambda}(\hat{\beta})\bigr)$, under moment restrictions, and the (unrestricted) empirical probability counterparts $\pi^{(0)} = \bigl\{\pi_i^{(0)}\bigr\}_{i=1}^n$, with $\pi_i^{(0)}=1/n$, for $i=1$, $2$, \ldots, $n$, that will assign equal probability weights for the sample moment restrictions, i.e., 
 \begin{align*}
(1/n)\sum\nolimits_{i=1}^n g\bigl(z(x_i;  \theta, \phi), \hat{\beta}\bigr) = 0.
\end{align*}
Next, define the {\it Kullback-Leibler divergence} $D_{\rm KL}$ between the (unrestricted) empirical probabilities $\pi^{(0)}$ and the implied empirical probabilities $\hat{\pi} \bigl(\hat{\beta}, \hat{\lambda}(\hat{\beta})\bigr)$, under moment restrictions, as follows
\begin{align}
D_{\rm KL} \left(\pi^{(0)} \bigl \Vert \hat{\pi} \bigl(\hat{\beta}, \hat{\lambda}(\hat{\beta})\bigr) \right) &= E_{\pi^{(0)}} \left( \log  \left( \pi^{(0)} \bigl /\hat{\pi} \bigl(\hat{\beta}, \hat{\lambda}(\hat{\beta})\bigr) \right) \right) \notag \\
&= \sum\nolimits_{i=1}^n \pi_{i}^{(0)} \log \left (\pi_{i}^{(0)} \bigl/ \hat{\pi}_i \bigl(\hat{\beta}, \hat{\lambda}(\hat{\beta})\bigr) \right) \notag \\
&= -(1/n) \sum\nolimits_{i=1}^n \log \left ( n \hat{\pi}_i \bigl(\hat{\beta}, \hat{\lambda}(\hat{\beta})\bigr) \right) \notag \\
&= - H \bigl(\pi^{(0)} \bigr) - (1/n) \ell \bigl(\hat{\beta}, \hat{\lambda}(\hat{\beta})\bigr), \label{Eq14}
\end{align}
where $\ell\bigl(\hat{\beta}, \hat{\lambda}(\hat{\beta})\bigr)$ is the empirical log-likelihood function, under moment restrictions, whereas the term $H \bigl(\pi^{(0)} \bigr)$ is the {\it Shannon entropy} corresponding to the (unrestricted) empirical probabilities $\pi^{(0)}$ satisfying
\begin{align}
0 < H \bigl(\pi^{(0)} \bigr) &= \sum\nolimits_{i=1}^n \pi_{i}^{(0)} \log \bigl(1\bigl/ \pi_{i}^{(0)}\bigr) = \log n \notag \\
                                        &\le \sum\nolimits_{i=1}^n \pi_{i}^{(0)} \log \bigl(1\bigl/ \hat{\pi}_i \bigl(\hat{\beta}, \hat{\lambda}(\hat{\beta})\bigr) \notag \\
                                        &= \sum\nolimits_{i=1}^n (1/n) \log \bigl(1 \bigl/ \hat{\pi}_i \bigl(\hat{\beta}, \hat{\lambda}(\hat{\beta})\bigr) \notag \\
                                        &= -(1/n) \ell\bigl(\hat{\beta}, \hat{\lambda}(\hat{\beta})\bigr), \label{Eq16}
\end{align}
for all $(\theta, \phi) \in \Theta \times \Phi$. Then, we have the following classical information inequality
\begin{align*}
D_{\rm KL} \left(\pi^{(0)} \bigl \Vert \hat{\pi} \bigl(\hat{\beta}, \hat{\lambda}(\hat{\beta})\bigr) \right) \ge \frac{1}{2}  \biggl \Vert \pi^{(0)}, \hat{\pi} \bigl(\hat{\beta} - \hat{\lambda}(\hat{\beta})\bigr) \biggr\Vert_{\rm TV}^2,
\end{align*}
i.e., the {\it Pinsker--Csisz\'{a}r--Kullback inequality}, that relates the total variational distance $\Vert \cdot \Vert_{\rm TV}$ and the {\it Kullback-Leibler divergence} $D_{\rm KL}$ (e.g., see \cite{r19} and \cite{r20} for additional discussions). Note that, from Equation~\eqref{Eq14}, we will have the following relationship
\begin{align*}
D_{\rm KL} \left(\pi^{(0)} \bigl \Vert \hat{\pi} \bigl(\hat{\beta}, \hat{\lambda}(\hat{\beta})\bigr) \right) + H \bigl(\pi^{(0)} \bigr) &= -(1/n) \ell\bigl(\hat{\beta}, \hat{\lambda}(\hat{\beta})\bigr).
\end{align*}
Moreover, taking aim at minimizing the negative of the empirical log-likelihood function $\ell\bigl(\hat{\beta}, \hat{\lambda}(\hat{\beta})\bigr)$, under moment restrictions, is equivalent to minimizing the total variational distance $\bigl\Vert \pi^{(0)} - \hat{\pi} \bigl(\hat{\beta}, \hat{\lambda}(\hat{\beta})\bigr) \bigr \Vert_{\rm TV}$, i.e., the gap on the discrepancy between the (unrestricted) empirical probabilities $\pi^{(0)}$ and that of the implied empirical probabilities $\hat{\pi} \bigl(\hat{\beta}, \hat{\lambda}(\hat{\beta})\bigr)$, under moment restrictions. Note that the quantity $H^{(0)}\bigl(\pi^{(0)}\bigr)$, which depends only on the observed data $y^{\rm obs}(x)$, is constant and evidently irrelevant in the subsequent optimization steps.

\end{document}